\documentclass[nohyper,12pt,letterpaper]{JHEP3}
\usepackage[all]{xy}
\usepackage{epsfig}
\usepackage{subfigure}
\usepackage{amsmath}
\usepackage{amsfonts}

\title{Real-time Fermions for Baryogenesis Simulations}
\author{
  Paul M. Saffin,\\
  School of Physics and Astronomy, University Park, University of
  Nottingham,\\ Nottingham NG7 2RD, UK\\
  Email: \email{paul.saffin@nottingham.ac.uk}}
\author{
  Anders Tranberg,\\
  Niels Bohr International Academy, Niels Bohr Institute,\\
  Blegdamsvej 17, 2200 Copenhagen, Denmark\\
  Email: \email{anders.tranberg@nbi.dk}}

\def\half{\textstyle{1\over2}}

\newcommand{\be}{\begin{equation}}
  \newcommand{\ee}{\end{equation}}
\newcommand{\ba}{\begin{eqnarray}}
  \newcommand{\ea}{\end{eqnarray}}
\newcommand{\bml}{\begin{mathletters}}
  \newcommand{\eml}{\end{mathletters}}

\newcommand{\del}{\ensuremath{\partial}}

\newcommand{\lab}{\label}

\abstract{We study how to numerically simulate quantum fermions out of thermal equilibrium, in the context of electroweak baryogenesis. We find that by combining the lattice implementation of Aarts and Smit \cite{Aarts:1998td} with the ``low cost" fermions of Borsanyi and Hindmarsh \cite{Borsanyi:2008eu}, we are able to describe the dynamics of a classical bosonic system coupled to quantum fermions, that correctly reproduces anomalous baryon number violation. To demonstrate the method, we apply it to the 1+1 dimensional axial $U(1)$ model, and perform simulations of a fast symmetry breaking transition. Compared to solving all the quantum mode equations as in \cite{Aarts:1998td}, we find that this statistical approach may lead to a significant gain in computational time, when applied to 3+1 dimensional physics. }

\preprint{}

\begin{document}

\section{Introduction\label{sec:intro}}

In electroweak baryogenesis \cite{Kuzmin:1985mm} the source of baryon number non-conservation is the quantum anomaly of fermions chirally coupled to the Standard Model SU(2) gauge field. When the gauge field evolves in such a way that its Chern-Simons number changes, the fermion, and hence B(aryon) and L(epton), number changes as
\ba
B(t)-B(0)=L(t)-L(0)=n_f[N_{\rm cs}(t)-N_{\rm cs}(0)],
\ea
where $n_f=3$ is the number of fermion generations in the Standard Model.
The question of successful baryogenesis thus reduces to whether a permanent change of Chern-Simons number can take place in the early Universe, presumably under the influence of CP-violation and the back-reaction of the fermions. 

Various models of baryogenesis have been proposed, of which the most popular (and most developed) is ``hot'' electroweak baryogenesis \cite{Cohen:1993nk}, where walls of bubbles nucleated in a first order phase transition interact in a CP-violating manner with the fermions in the hot plasma. In this way a net left-right fermion asymmetry is generated inside and outside the bubbles, and equilibrium gauge dynamics (sphaleron transitions) convert this asymmetry into a baryon asymmetry.

The rate of sphaleron transitions can reliably be calculated in thermal equilibrium using sophisticated Monte-Carlo methods \cite{Moore:1999fs,D'Onofrio:2010es}. In such a setup, fermions can be included in terms of effective couplings for the bosonic theory, for instance through dimensional reduction \cite{Kajantie:1995dw}.

An alternative scenario is ``Cold'' electroweak baryogenesis \cite{GarciaBellido:1999sv,Krauss:1999ng,Rajantie:2000nj,Copeland:2001qw,Tranberg:2003gi,Tranberg:2009de}, where the electroweak phase transition does not involve bubble nucleation, but instead a fast quench of the Higgs potential. Here, baryon number violating processes are not equilibrium Sphaleron transitions, but complicated out-of-equilibrium field dynamics.

Numerical real-time simulations of electroweak baryogenesis have until now neglected dynamical fermions. Instead, purely bosonic systems are evolved and baryon number has simply been assumed to follow the gauge field Chern-Simons number in accordance with the anomaly equation, ignoring fermionic backreaction. 

One case where this is certainly not allowed is for minimal electroweak baryogenesis, since CP-violation in the Standard Model originates from the fermion mass matrix. A possible approach employed in \cite{Shaposhnikov:1987tw,GarciaBellido:1999sv,Tranberg:2003gi} is to integrate out the fermions in the path integral or in perturbation theory, thus recovering CP-violation effects in terms of a series of higher-dimensional bosonic terms.

The current understanding that Standard Model CP-violation is strongly suppressed at high temperatures, and therefore insufficient for successful baryogenesis follows from such a computation (see for instance \cite{Shaposhnikov:1987tw,Rubakov:1996vz}). In contrast, at low temperatures relevant for ``Cold'' baryogenesis, recent calculations have shown that the suppression is absent \cite{Smit:2004kh,Hernandez:2008db,GarciaRecio:2009zp,Salcedo:2011hy}, and direct  numerical simulations have in turn indicated that Standard Model CP-violation may in fact be large enough to accommodate the observed asymmetry \cite{Tranberg:2009de,Tranberg:2010af}. 

A possible caveat to this procedure is that it is based on a gradient expansion in the gauge and Higgs fields, which may not be valid during electroweak symmetry breaking. And so although the work in \cite{Tranberg:2009de,Tranberg:2010af} is very encouraging indeed, it would be even better not having to integrate out the fermions, but include them directly in real-time simulations of the transition. In this way, the CKM matrix and CP-violation could be included from first principles.

In \cite{Aarts:1998td} Aarts and Smit showed how to implement quantum fermions in real-time, coupled to classical bosonic gauge and scalar fields. The method involves a proper lattice discretization in Minkowski space, and the realisation that since fermions are bilinear in the action, the field operators can be expanded into mode functions, in terms of time-independent creation-annihilation operators. These mode functions can then be solved in the classical bosonic background, with the back-reaction on the bosonic fields defined as the quantum averages over the creation-annihilation operators for some given initial state. 

In practice, the problem is that for every momentum mode $k$ (equal to the number of lattice sites $n_x^D$, where $D$ is the number of spatial dimensions), one needs to solve a separate real-time field equation (the mode function equation) for which the numerical effort is also proportional to the number of lattice sites. Hence the total numerical problem scales as $n_x^{2D}$, and quickly becomes unmanageable for large three-dimensional lattices. Large lattices are often required in baryogenesis simulations to accommodate extended objects such as sphalerons and for having enough infrared modes for a fast quench to be correctly reproduced.

Some time ago \cite{Borsanyi:2008eu}, Borsanyi and Hindmarsh showed how to replace the $n_x^D$ mode equations by an ensemble of fermion field realisations, approximating the quantum fermion expectation values through a statistical averaging procedure. In the context of a scalar-fermion theory, they showed that one can significantly reduce the numerical effort, at least in three dimensions. This is because the number of random realisations in the ensemble $N_q$ can be much smaller than $n_x^D$. 

In this work, we will implement the ``low cost fermion'' or ``fermion ensemble'' method of Borsanyi and Hindmarsh to the 1+1 dimensional axial-U(1) model with fermions of Aarts and Smit. This will act as a toy model for the electroweak part of the Standard Model, and will provide a testing ground for the method. In particular, we will investigate whether this method correctly reproduces the anomaly equation, charge conservation and the correct dynamics, and determine how large the fermion ensemble needs to be to get reliable results. We also want to understand when it is correct to neglect fermion backreaction for the boson dynamics.

The paper is structured as follows: In section \ref{sec:setup}, we will introduce the model, discretize it on the lattice (section \ref{app:lattice}), and derive the equations of motion. In section \ref{sec:bosonfermion} we introduce an adapted version of the ``Male'' and ``Female'' fermion fields \cite{Borsanyi:2008eu} required to generate the fermion correlators with c-number fields.  In section \ref{sec:results} we describe the numerical setup and the results, and we conclude in section \ref{sec:conc}.

\section{The Axial-U(1)-Higgs-fermion model in 1+1 dimensions \label{sec:setup}}

We will consider the 1+1 dimensional Abelian-Higgs model, coupled axially to fermions. The action reads in the continuum:
\ba
\label{eq:action}
S=S_H+S_A+S_{F},
\ea
in terms of the components
\ba
S_H&=&-\int\;d^2x\;\left[D_\mu\phi^\dagger D^\mu\phi+\lambda(\phi^\dagger\phi-v^2/2)^2\right],\\
S_A&=&-\int\;d^2x\;\frac{1}{4e^2}F_{\mu\nu}F^{\mu\nu},\\
S_F&=&-\int\; d^2x\;\left[\bar{\psi}\gamma^\mu\left(\partial_\mu+iA_\mu\gamma_5\right)\psi+G\bar{\psi}\left(\phi^*P_L+\phi P_R\right)\psi\right],
\ea
and with the definitions
\ba
D_\mu\phi=\del_\mu\phi-iA_\mu\phi,&\quad& F_{\mu\nu}=\del_\mu A_\nu-\del_\nu A_\mu,\quad P_{R,L}=\frac{1}{2}\left(1\pm\gamma_5\right).
\ea
The action is invariant under gauge transformations of the form
\ba
\psi\rightarrow\exp(iq\xi(x)\gamma_5)\psi,\quad \phi\rightarrow\exp(-i\xi(x))\phi,\quad A_\mu\rightarrow A_\mu-\del_\mu\xi(x),
\ea
if we take $q=1/2$. 

In lattice simulations it is more convenient to work with vector gauge symmetry, rather than axial, and so noting that the left and right chiral components have opposite charge, it is therefore natural to charge-conjugate one of them \cite{Aarts:1998td},
\ba
\psi_R=(\bar{\psi}_R'\mathcal{C})^T,\quad \bar{\psi}_R=-(\mathcal{C}^\dagger\psi_R')^T,\quad \psi_L=\psi_L',\quad \bar{\psi}_L=\bar{\psi}_L',
\ea
where $\mathcal{C}$ is the charge-conjugation matrix given in Appendix \ref{app:cons}. Upon doing this, the action in the new variables (but omitting the primes) reads
\ba
S_F=-\int\; d^2x\;\left[\bar{\psi}\gamma^\mu\left(\partial_\mu-iqA_\mu\right)\psi+\half G\psi^T\mathcal{C}^\dagger\phi^*\psi-\half G \bar{\psi}\mathcal{C}\phi\bar{\psi}^T\right].
\ea
It is no longer axially coupled, but the Yukawa interaction has become a Majorana term, and the gauge symmetry has become vector-like
\ba
\psi\rightarrow\exp(-iq\xi(x))\psi,\quad
\phi\rightarrow\exp(-i\xi(x))\phi,\quad
A_\mu\rightarrow A_\mu-\del_\mu\xi(x),
\ea
with the continuum equations of motion being
\ba
\label{eq:higgs_eom}D_\mu D^\mu\phi-2\lambda(\phi^\star\phi-v^2/2)\phi-\frac{G}{2}\psi^TC\psi&=&0,\\
\label{eq:ferm_eom}\gamma^\mu D_\mu\psi+G\phi\psi^\star&=&0,\\
\label{eq:gauge_eom}\del_\mu F^{\mu\nu}+e^2(j_{(\psi)}^\nu+j_{(\phi)}^\nu)&=&0.
\ea
We have introduced the gauge currents
\ba
\label{eq:fermiCurrents}
j^\mu_{(\psi)}=iq\bar{\psi}\gamma^\mu\psi,
\quad j^\mu_{(\phi)}=i(\phi D^\mu \phi^\star-\phi^\star D^\mu\phi).
\ea
There is one further symmetry of this system, the one that this work is principally interested in, and it is the global U(1) symmetry, $\psi\rightarrow\exp(-i\omega\gamma^5)\psi$. This symmetry has an associated current
\ba
j^\mu_5=i\bar{\psi}\gamma^\mu\gamma^5\psi,
\ea
which is precisely the fermion current in the original\footnote{Non-charge conjugated.} theory, and classically conserved if one naively applies the equations of motion. Quantum mechanically, however, it is the subject of an anomaly
\ba
\del_\mu j^\mu_5&=&\frac{1}{4\pi}\epsilon^{\mu\nu}F_{\mu\nu}=\del_\mu C^\mu,\quad C^\mu=\frac{1}{2\pi}\epsilon^{\mu\nu}C_\nu,
\ea
and this allows us to relate the total fermion number, $Q(t)=\int dx\;j^0_5$, to the Chern-Simons number, $C(t)=\int dx\;C^0=-\frac{1}{2\pi}\int dx A_1(x)$, through
\ba
Q(t_f)-Q(t_i)=C(t_f)-C(t_i).
\ea

There is one further number that is worth mentioning, the winding number of the Higgs field. When the Higgs field is away from zero, it takes values on a circle parametrized by its phase $\theta$, $\phi(x)=|\phi(x)|e^{i\theta(x)}$. Using this phase we may define a Higgs winding number, describing the number of times the field winds around this circle on a given spatial section,
\ba
N_W=\frac{1}{2\pi}\int dx\, \partial_1\theta(x).
\ea
In a vacuum state we know that the covariant derivative of the Higgs field vanishes, and that its modulus is constant, in which case we have that $\del_x\theta=A_x$, leading to the sum of the Higgs winding and Chern-Simons numbers vanishing in the vacuum.

For the numerical work, we discretize the Abelian-Higgs-fermion model on a 1+1 dimensional lattice of size $L=a_1 n_x$ at the level of the action, and derive lattice equations of motion as described in Appendix \ref{app:lattice}.

\section{Bosons and fermions\label{sec:bosonfermion}}

We are interested in the time-evolution of this system, and we will adopt the approach of \cite{Aarts:1998td}, where the dynamics of bosonic and fermionic degrees of freedom are treated differently. The gauge and scalar fields are evolved using the classical equations of motion described in the previous section. Classical dynamics is an excellent approximation to the quantum dynamics for processes dominated by infrared physics and for fields with large occupation numbers. The fermions are treated completely quantum-mechanically, in the sense of solving the quantum equation of motion (\ref{eq:ferm_eom}) in the classical bosonic background, in terms of field operators. Since the fermions are bi-linear in the action, the equation of motion is linear, and the field can in all generality be expanded in terms of a set of mode functions and time-independent creation-annihilation operators (see below). 

This leaves the question of the back-reaction of the fermions on the classical bosonic fields. Following \cite{Aarts:1998td} again, we interpret the fermionic terms in the gauge and scalar equations of motion as expectation values of the corresponding operators, evaluated in some state encoded in the expectation values of the creation-annihilation operators. These states are time-independent, and amount to specifying an initial condition. The time-evolution is in the mode functions only.

We then take one step further by representing these creation-annihilation operators by a set of random numbers, thereby generating an ensemble of fermion field-realisation \cite{Borsanyi:2008eu}. These can each be evolved in the same bosonic background, and the field expectation values are then replaced by simple averages over the ensemble. The point is to note that the number of field realisations ($N_q$) in the ensemble can be much smaller than the number of mode functions ($n_x^D$), and the statistical approach can therefore be much cheaper in terms of computational effort.

\subsection{Boson initialisation\label{sec:bosoninit}}

We will consider two setups for the bosonic fields. The first (in section \ref{sec:handmadeG0}) is to by hand set the gauge-Higgs evolution to be a sequence of sphaleron transitions, thus forcing the Chern-Simons number to change (as in \cite{Aarts:1998td}). The fermion fields evolve dynamically in the background of these handmade sphalerons. We will use this setup to test the ability of the ensemble to capture the anomaly, and to find out how large the ensemble needs to be. 

When considering the non-perturbative field dynamics (in sections \ref{sec:fulldynamicsG0} and \ref{sec:CEB}), we instead initialise the bosonic fields by setting $A_\mu({\bf x},t=0)=0$, $\partial_0 \phi({\bf x},t=0)=0$ and introducing random noise for the scalar field, $\phi({\bf x},t=0)$ 
\ba
\phi(x)=\frac{1}{\sqrt{2}}\left(\phi_1(x)+i\phi_2(x)\right),\quad \phi_{1,2}(x)=\int \frac{dk}{2\pi}\phi^{1,2}_k e^{ikx},
\ea
in terms of random numbers $\phi_k^{1,2}$, with the correlator
\ba
\langle\phi_k^{1,2}{\phi_k^{1,2}}^*\rangle=\frac{\omega_k}{2},\quad \omega_k=\sqrt{k^2_{\rm lat}+\lambda v^2}.
\ea
The gauge field momenta $\partial_0 A_1({\bf x},t=0)$ are found by solving the Gauss constraint (\ref{eq:gauge_eom}) with fermion sources. 

As described in \cite{Smit:2002yg,GarciaBellido:2002aj} this initialisation represents\footnote{In fact, we should also initialise the momenta $\partial_0\phi$ with random numbers for the identification with the quantum vacuum to be completely correct. Setting $\partial_0\phi$ to zero initially makes the initial total charge on the lattice vanish, a requirement for consistency of Gauss law. To achieve this is cumbersome, but possible, when initialising both field and momenta. For our purposes here, initialising only the field variables will suffice.} an initial quantum vacuum before Higgs symmetry breaking, $V_{\rm ini}=\lambda v^2\phi^*\phi$. In the subsequent evolution, momentum modes $k^2<\lambda v^2$ will grow exponentially, and from some time on they can be described using classical dynamics. The fermions do not grow, and are still treated quantum mechanically. 

The amount of growth of the scalar modes is determined by the (in 1+1 dimensions dimensionless) parameter $v$. This can be seen in various ways. The growth lasts until backreaction from self-interactions kick in, i.e. when $\phi^2\simeq v^2$. For a given mode, we have
\ba
\langle\phi^*(x)\phi(x)\rangle=\int\frac{dk}{2\pi}\langle\phi_k^\dagger\phi_k\rangle=\int\frac{dk}{2\pi}\frac{n_k+\frac{1}{2}}{\omega_k},
\ea
where initially, $n_k=0$. Classical dynamics is a good approximation once the mode has grown so much that $n_k+1/2\gg 1/2$. Hence large $v$ allows for classicality. 

Another way of phrasing this is to note that once $\phi\simeq v$, the scalar-gauge interaction and the effect of the scalar on the fermions goes as $ev$ and $Gv$, respectively, whereas back-reaction of fermions on bosons is $e$ and $G$. Hence for large $v$, fermion effects are relatively smaller (the fields have relatively smaller amplitude).

In the following, we will employ $v=64$ and $v=8$. Since only modes with $k^2<\lambda v^2$ are unstable, only they will be classical, and these are therefore the only bosonic modes we initialise.

\subsection{Fermion mode expansion\label{sec:modeexpansion}}

Now we need to know how to set up the initial conditions for the fermion field, and for this we will be using the usual mode expansion. There is a slight complication, however, due to the fact that the fermion equation of motion is not linear in $\psi$, but involves both $\psi$ and $\psi^\star$ (\ref{eq:ferm_eom}). This leads to the real and imaginary components having different equations of motion, particularly when the lattice Wilson term is included, and so it is convenient to write the Dirac spinor as a combination of two Majorana spinors, which in our conventions (Appendix \ref{app:cons}) just means breaking $\psi$ into real and imaginary parts.
\ba
\psi=\frac{1}{\sqrt{2}}[\Psi_1-i\Psi_2],
\ea
and it is these components that we write as a mode expansion
\ba
\Psi(t,x)&=&\int\frac{dk}{2\pi}\frac{1}{2\omega_k}\left[b_kU_ke^{ik.x}+b^\dagger_kV_ke^{-ik.x}\right],
\ea
in terms of the constant spinors $U$ and $V$ (given in Appendix \ref{app:cons}) and a set of creation-annihilation operators $b^\dagger_k$, $b_k$. We then note that the fields $\Psi_{1,2}$ are canonically normalized, and that their conjugate momenta are $i\Psi_{1,2}^T$, so the canonical anti-commutations relations are
\ba
\left\{\Psi_\alpha(t,\underline x),\Psi_\beta(t,\underline x')\right\}&=&\delta(\underline x-\underline x')\delta_{\alpha\beta},
\ea
which may be achieved by imposing
\ba
\left\{b_k,b_{k'}^\dagger\right\}&=&(2\pi)2\omega_k\delta(k-k').
\ea
In the equations of motion for the bosonic fields we require the quantum expectation value of fermion bilinears, and so we follow \cite{Borsanyi:2008eu} in constructing the two-point functions
\ba
D^>_{\alpha\beta}(x,y)&=&\langle |\Psi_\alpha(x)\Psi_\beta(y)| \rangle,\quad
D^<_{\alpha\beta}(x,y)=-\langle |\Psi_\beta(y)\Psi_\alpha(x)| \rangle,\\
D_{\alpha\beta}(x,y)&=&\frac{1}{2}\left[D^>_{\alpha\beta}(x,y)+D^<_{\alpha\beta}(x,y)\right],
\ea
leading to
\ba
\label{eq:fermcorr}D_{\alpha\beta}(x,y)&=&\frac{1}{2}\sum\frac{dk}{2\pi}\frac{1}{2\omega_k}\left[U_{k\alpha}V_{k\beta}e^{ik.(x-y)}-V_{k\alpha}U_{k\beta}e^{-ik.(x-y)}\right].
\ea
where we take $b_k| \rangle=0$. We note that although the fields are real, the two-point function is imaginary, $D^\star_{\alpha\beta}(x,y)=-D_{\alpha\beta}(x,y)$. The observation of \cite{Borsanyi:2008eu} is that we can construct a bi-linear of classical spinor fields, for which the ensemble average two-point function matches (\ref{eq:fermcorr}). This allows us to simulate the quantum backreaction of fermion fields using ensemble averages of classical spinor fields; this is what we shall now do.

\subsection{Fermion ensemble, Male and Female\label{sec:malefemale}}

If we were to simply evolve an ensemble of fermions, where we draw the initial conditions of each realization from a sample with the appropriate distribution and then take the ensemble average $\langle \Psi(x)\Psi(y)\rangle$ to mimic the quantum two-point function, we cannot reproduce (\ref{eq:fermcorr}). However, if one introduces two "genders" of fermions, male and female, and writes their mode expansion as
\ba
\Psi_M(x)&=&\frac{1}{\sqrt 2}\int\frac{dk}{2\pi}\frac{1}{2\omega_k}\left[\eta_{k}U_{k}e^{ik.x}+\eta^\star_{k}V_{k}e^{-ik.x}\right],\\
\Psi_F(x)&=&\frac{i}{\sqrt 2}\int\frac{dk}{2\pi}\frac{1}{2\omega_k}\left[\eta_{k}U_{k}e^{ik.x}-\eta^\star_{k}V_{k}e^{-ik.x}\right],
\ea
then we find that taking
\ba
\langle\eta_k\eta^\star_p\rangle&=&(2\pi)2\omega_k\delta(\underline k-\underline p),\quad\langle\eta_k\eta_p\rangle=0,
\ea
leads to
\ba
i\langle\Psi_{M\alpha}\Psi'_{F\beta}\rangle&=&\frac{1}{2}\sum_k\frac{dk}{2\pi}\frac{1}{2\omega_k}\left[U_{k\alpha}V_{k\beta}e^{ik.(x-y)}- V_{k\alpha}U_{k\beta}e^{-ik.(x-y)}\right]\,\\
\label{eq:correlatorEquiv}
 &=&D_{\alpha\beta}(x,y),
\ea
so we now have an explicit way of replacing quantum averages, $\langle |X|\rangle$, with ensemble averages, $\langle X\rangle$. This leads to us evolving
\ba
D_\mu D'^\mu\phi-2\lambda(\phi^\star\phi-v^2/2)\phi-\frac{iG}{2}\langle\psi^{M,T}C\psi^F\rangle&=&0,\\
\gamma^\mu \tilde D_\mu\psi^{M,F}+G\phi\psi^{M,F,\star}&=&0,\\
\del_\mu (\del'^\mu A^\nu-\del'^\nu A^\mu)+e^2(j_f^\nu+j_b^\nu)&=&0,
\ea
rather than the equations of motion appearing in Appendix \ref{app:lattice}. The fermion gauge-current is also modified in this prescription, with the requirement of its conservation leading to
\ba
j_{f,\mu}&=&\frac{iq}{4}\left[i\bar\psi^M(x)\gamma_\mu U^q_\mu(x)\psi^F(x+\mu)+i\bar\psi^M(x+\mu)\gamma_\mu U^{q\star}_\mu(x)\psi^F(x)\right.\nonumber\\
   &~& \left.-i\bar\psi^F(x+\mu)\gamma_\mu U^{q\star}_\mu(x)\psi^M(x)-i\bar\psi^F(x)\gamma_\mu U^q_\mu(x)\psi^M(x+\mu)\right].
\ea
Furthermore, we need a representative of the anomalous current,
\ba
j_{\mu,5}&=&\frac{i}{4}\left[ i\bar\psi^M(x)\gamma_\mu\gamma_5U_\mu(x)\psi^F(x+\mu)+i\bar\psi^M(x+\mu)\gamma_\mu\gamma_5U_\mu^\star(x)\psi^F(x)\right.\nonumber\\
   &~&    \left.-i\bar\psi^F(x+\mu)\gamma_\mu\gamma_5U^\star_\mu(x)\psi^M(x)-i\bar\psi^F(x)\gamma_\mu\gamma_5U_\mu(x)\psi^M(x+\mu)\right].
\ea
Because of cancellation between lattice doublers, this quantity is conserved for vanishing Wilson term ($r_1=0$, see Appendix \ref{app:lattice}), but for $r_1=1$ the current is anomalous, as we will see below. 

\section{Results \label{sec:results}}

\subsection{Hand-made Sphaleron transitions\lab{sec:handmadeG0}}

\DOUBLEFIGURE{./Pictures/Presph128.eps,width=7cm,clip}{./Pictures/nxsph.eps,width=7cm,clip}
{The bosonic fields are evolved through a series of sphaleron transitions, thereby continuously changing the Chern-Simons number. The fermion fields are evolved in this background using the equations of motion, and fermion number is seen to obey the anomaly equation. The Higgs winding number follows in steps (shown is $-N_{\rm w}$).\label{fig:sample_sph} }{For large enough Chern-Simons number, the anomaly equation is no longer satisfied on a small lattice. Increasing the volume allows a larger range of agreement. The three curves coincide upon rescaling by the lattice size $n_x$ along both axes (insert).\label{fig:anfail} }

We first want to check that our approach of replacing mode functions by a random ensemble still leads to correct dynamics and that the anomaly equation is satisfied, as in \cite{Aarts:1998td}. We also want to determine how large the ensemble needs to be to get statistically reliable results for the anomaly and the dynamics.

An elegant way of doing this is to by hand set the gauge-Higgs field evolution to be a series of sphaleron transitions, thereby continuously changing the Chern-Simons number in a controlled way. The explicit expression for the bosonic fields can be found in \cite{Aarts:1998td}. The important point is that sphaleron transitions take place at half-integer values of $t/t_0$, and the fields are in vacuum at integer values. We choose the timescale $t_0$ so that the transitions are slow enough that the fermions, which are evolved using the equations of motion in the sphaleron-vacuum background, do not lag too much behind, $m_A t_0=4$. From the point of view of the fermion, the evolution is almost adiabatic, and no additional spurious particle creation takes place. Only the particles associated with the anomaly contribute. At the sphaleron configuration, the Higgs field length vanishes, and Higgs winding changes discontinuously from one integer to the next. In the vacuum, $N_{\rm w}=-N_{\rm cs}$, and we will always plot $-N_{\rm w}$.

Fig.~\ref{fig:sample_sph} shows the evolution of Chern-Simons number $N_{\rm cs}$, Higgs winding number $-N_{\rm W}$ and the fermion number $N_{\rm f}$. The parameters used were $m_AL=evL=25.6$, $G=0$, $N_q=90$, $n_x=128$, $v=2$, timestep $dt=0.05$. The agreement between Chern-Simons number and fermion number is remarkably precise, even for such a small ensemble. 

As was pointed out in \cite{Aarts:1998td}, fermion number is periodic on a finite lattice with period $2n_x$, and so for large $N_{\rm cs}$ the agreement will fail. Fig.~\ref{fig:anfail} shows fermion number for very large Chern-Simons number at different values of the lattice size $n_x$ (volume is fixed $evL=6.4$). These show the lattice behaviour, and can indeed be rescaled by $n_x$ (along both axes) to end up on top of each other (inset). This is exactly as in \cite{Aarts:1998td}, and means that sufficiently large lattices can accommodate any Chern-Simons number. Notice that the ensemble is still $N_q=90$. 

\DOUBLEFIGURE{./Pictures/Nsph.eps,width=7cm,clip}{./Pictures/Nsph128.eps,width=7cm,clip}
{Convergence of the fermion number as the ensemble is enlarged. Here 10 to 2430 realisations, for a small lattice $n_x=32$. \label{fig:Ndep} }{Convergence of the fermion number as the ensemble is enlarged. Here 10 to 2430 realisations, for a large lattice $n_x=128$.\label{fig:Ndep128} }

Although the small ensemble very convincingly reproduces the anomaly when looked at by eye, it is only prudent to investigate the statistical precision. This is shown in Figs.~\ref{fig:Ndep} and \ref{fig:Ndep128} for ensembles of 10, 30, 90, 270, 810 and 2430 random realisations, respectively. The left-hand plot is on a $n_x=32$ lattice, and we see that the agreement is always fairly good, even for $N_q=10$. Looking closer (inset), we do see that the curves converge, and in fact converge to a value slightly off $N_{\rm cs}$. This is the finite volume effect as described before. In the right-hand plot with $n_x=128$, this discrepancy is gone and increasing the ensemble, fermion number converges to the Chern-Simons number value. We conclude that convergence in $N_q$ is achieved at the few-percent level for $N_q=\mathcal{O}(1000)$.

\DOUBLEFIGURE{./Pictures/Gsph.eps,width=7cm,clip}{./Pictures/nxsph3.eps,width=7cm,clip}
{Fermion number for a small lattice $n_x=32$, with increasing values of the Yukawa coupling $G$. Chern-Simons number is shown for comparison. Lattice artifacts are larger for non-zero $G$. \label{fig:Gsph1} }{Fermion number for fixed $G=0.1$ and volume $evL$, but increasing $n_x$, decreasing lattice spacing. Also for finite Yukawa coupling, lattice artifacts can be removed by increasing the number of lattice points or increasing the volume. \label{fig:Gsph2} }

As reported in $\cite{Aarts:1998td}$, including the Yukawa coupling $G$ makes the lattice artefacts stronger. Fig.~\ref{fig:Gsph1} shows the fermion number in the hand-made sphaleron background for $n_x=32$, $evL=6.4$ with varying $G$. The anomaly holds until $N_{\rm cs}\simeq 5$, after which the finite size effects kick in, stronger with increasing $G$. However, increasing the lattice size again ameliorates the situation, as shown in Fig.~\ref{fig:Gsph2}, where $G/e=0.1$ is kept constant, and the lattice discretization is made finer (constant volume $evL=6.4$ and $n_x$ increasing\footnote{Increasing $n_x$ with increasing physical volume has the same effect.}).

\EPSFIGURE{./Pictures/LargeVsph.eps,width=7cm,clip}
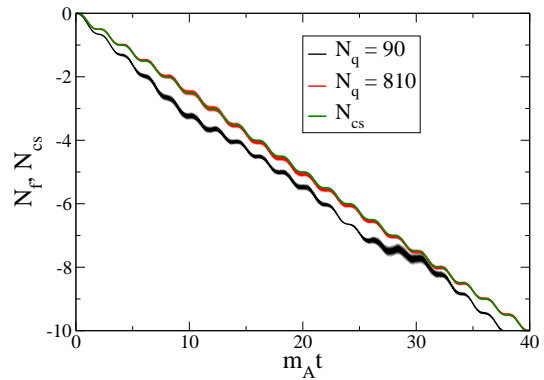
{Sphaleron transitions and fermion number for larger volume $evL=51.2$, at $G/e=0.1$. Larger statistics cures the discrepancy. \label{fig:largeV} }

From a practical point of view, we would like to be able to run baryogenesis simulations for timescales $m_At=evt=\mathcal{O}(100)$, on a large enough lattice to fit in the appropriate physics $m_AL=evL\gg 1$, while having the anomaly correctly reproduced for realistic values of the Chern-Simons number, say $N_{\rm cs}=\mathcal{O}(10)$. And we also need to include a non-zero Yukawa coupling, at least for physics around the electroweak transition. The question is whether we can find a combination of $evL$, $n_x$, $G$ and $N_q$ that can accommodate this.

Fig.~\ref{fig:largeV} shows a run on a much larger lattice, $evL=51.2$, $v=2$, with Yukawa coupling $G/e=0.1$, in a range of $N_{\rm cs}=0-10$. We first note that the larger volume makes the anomaly agree less well than for the runs in Fig.~\ref{fig:Gsph2}, which did reasonably well until $N_{\rm cs}=5$. This is the case both for $n_x=256$ and for $n_x=512$, with half the lattice spacing (not shown). However, this is just adding up of statistical fluctuations, and can be compensated for by increasing the ensemble size. At $N_q=810$ the agreement is again convincingly reproduced. The finite volume effect is not apparent at these lattice sizes.

\subsection{Non-equilibrium dynamics\lab{sec:fulldynamicsG0}}

\DOUBLEFIGURE{./Pictures/noneqG0.eps,width=7cm,clip}{./Pictures/noneq3.eps,width=7cm,clip}
{The Higgs field, Chern-Simons number, winding number and fermion number during a tachyonic transition. Chern-Simons number and fermion number are indistinguishable.\label{fig:noneqG0} }{Convergence in $N_q$ of Chern-Simons and fermion number for $v=64$. Convergence is excellent until $m_At\simeq 100$, after which the effects of statistical fluctuations in the fermions have accumulated enough to make a difference.\label{fig:noneq3} }

In a fast-quench symmetry breaking transition, Higgs field modes with $k^2<\lambda v^2$ will be unstable (``tachyonic'') and grow exponentially. This drives the gauge field to also grow until non-linear backreaction begins to dominate, stop the growth and eventually leads to thermalisation. In the presence of CP-violation, such a transition may lead to a baryon asymmetry (see for instance \cite{Tranberg:2009de}).

For our choice of initial conditions, the initial gauge field is driven by the fermion ensemble fluctuations and the initial scalar field. Our goal is that for a given scalar field configuration, the evolution should be independent of $N_q$, so that the statistics reliably reproduce the fermion state. We need a large enough ensemble to have statistical fluctuations under control. We will set
\ba
\frac{\lambda}{e^2}=\frac{1}{4},\quad n_x=256, \quad m_A=ev=0.2, \quad m_H=\sqrt{2\lambda v^2},
\ea
and vary $N_q$.

Fig.~\ref{fig:noneqG0} shows the Higgs field (black line), Higgs winding number (green), Chern-Simons number (red) and fermion number (blue) in such a transition. $G=0$, $v=64$ and $N_q=2430$. The Higgs field ``falls off the hill'' as expected, and performs oscillation around its finite temperature minimum. Meanwhile, the Chern-Simons number grows and oscillates near the integer-value Higgs winding number. The anomaly is so well obeyed that fermion number is essentially indistinguishable from Chern-Simons number. The Higgs winding bounces around in the beginning, an effect of the Higgs field length being small, and the winding number therefore ill-defined. But once symmetry breaking gets going, winding number is stable, integer and consistent with the Chern-Simons number.

We illustrate the convergence of the dynamics with increasing $N_q$ in Fig.~\ref{fig:noneq3}, where tachyonic transitions are performed for $v=64$ for different sizes of the ensemble. As expected, we see convergence in $N_q$, but also that the required ensemble is $\mathcal{O}(1000)$, to get agreement at this value of $v$ and for these times. In fact, the Chern-Simons number is very sensitive to fluctuations in the fermion source. This is at least partly because in 1+1 dimensions, a U(1) gauge field only has one dynamical degree of freedom (i.e. up to gauge transformations), which is precisely the Chern-Simons number. This means that in the sea of fermion degrees of freedom, the single degree-of-freedom gauge field can easily be bounced around. These issues are specific for 1+1 dimensions, and we will proceed with $v=64$ and $N_q=2430$, for which convergence is under control at least for $m_A t<100$, and qualitatively correct for $m_A t<150$. This will suffice for the present work, but can be improved depending on the level of precision required.

Since the Yukawa coupling is absent, the fermions are massless throughout. Also, the fermion and boson total charges are individually conserved at the level of $\mathcal{O}(10^{-13})$, and Gauss law is conserved at a (relative) level of $\mathcal{O}(10^{-8})$. 

\DOUBLEFIGURE{./Pictures/noneqG2.eps,width=7cm,clip}{./Pictures/noneqG1.eps,width=7cm,clip}
{Chern-Simons number, Higgs winding number, fermion number and Higgs field in a tachyonic transition at $G/e=0.1$, $N_q=2430$, $v=64$. \label{fig:noneqG1} }{Convergence in $N_q$ of Chern-Simons and fermion number for $v=64$ at $G/e=0.1$. At finite Yukawa coupling, we need a somewhat larger ensemble to reach convergence, here $N_q = 2430-7290$.\label{fig:noneqG2} }

For non-zero Yukawa coupling the fermions acquire masses as the Higgs transition proceeds (in addition to the gauge and Higgs fields). As we saw in section \ref{sec:handmadeG0}, the Yukawa coupling introduces stronger lattice artefact, but these could be cured by using larger ensembles. In Fig.~\ref{fig:noneqG1} and \ref{fig:noneqG2} we show the evolution and convergence in $N_q$ of a simulation with $G/e=0.1$, $v=64$. We see that although there is a clear effect of non-zero $G$, convergence still holds by increasing the ensemble to a few thousand, and fermion number (dashed lines) follows Chern-Simons number fairly well. 

\subsection{Application: Cold Electroweak Baryogenesis in 1+1 dimensions\label{sec:CEB}}

\DOUBLEFIGURE{./Pictures/noneqkappa.eps,width=7cm,clip}{./Pictures/noneqkappa2.eps,width=7cm,clip}
{The ensemble averages of Higgs field, Chern-Simons number, fermion number and Higgs winding number over 64+64$^*$ scalar realisations, with $v=64$ and $\kappa=0.04$.\label{fig:noneqCP1} }{A zoom-in of Fig.~12. The asymmetry is driven by the oscillation of the Higgs field, through the C(P) violating force term. Winding number can only change when the (local) Higgs field is small.\label{fig:noneqCP2} }

In the minimal model of electroweak baryogenesis, CP-violation is provided through the CKM fermion mass matrix. For hot baryogenesis, this effect is much too small to account for the asymmetry and a separate source of CP-violation is required. The situation is less clear for ``cold'' baryogenesis (see for instance \cite{Tranberg:2009de}). For illustration, we will postpone this issue, and simply introduce C(P) violation\footnote{From the point of view of the present model, we actually break C and P separately, while CP is conserved. This is the analogue of requiring CP violation in 3+1 dimensions. } in our 1+1 dimensional model through a bosonic term in the action (exactly as in \cite{Smit:2002yg}),
\ba
S\rightarrow S-\int d^2x\frac{\kappa n_f}{4\pi}\,\phi^*\phi\, \epsilon_{\mu\nu}F^{\mu\nu},
\ea
which amounts to an addition to the bosonic equations of motion (\ref{eq:gauge_eom}), (\ref{eq:higgs_eom}) of 
\ba
\partial_0^2A_1 &=& \ldots +\frac{e^2n_f\kappa}{2\pi}\partial_0 |\phi|^2,\\
\partial_0^2\phi &=& \ldots +\frac{n_f\kappa}{2\pi}\partial_0A_1\phi,\\
\partial_1\partial_0A_1&=&\ldots-\frac{n_f\kappa}{2\pi}\partial_1|\phi|^2.
\ea
The conservation of Gauss law and the anomaly and the convergence in $N_q$ is unaltered by this addition, and the C(P) violation is not obvious from a given random realisation of the bosonic fields. We now need to also average over an ensemble of bosonic realisations, each with a separate ensemble of fermion fields. This represents a quantum initial state of the Higgs fields coupled to fermions initially in the vacuum. 

Fig.~\ref{fig:noneqCP1} shows the scalar-ensemble averaged observables, Higgs field, Chern-Simons number, fermion number and winding number, at $\kappa=0.04$, $v=64$ and $N_q=2430$. We average over a set of 64 random realisations plus the corresponding C(P)-conjugate configurations. This makes the ensemble explicitly C(P) symmetric, and the asymmetry will be identically zero for $\kappa=0$. This procedure is similar to the one employed in \cite{Tranberg:2006dg,Tranberg:2006ip}. We see that an asymmetry is indeed generated in Chern-Simons, winding and fermion numbers as the transition proceeds. The anomaly is very well obeyed until times $m_At\simeq 100$, where fermion number goes a little low. We checked that this is indeed due to the configurations with relatively large $N_{\rm cs}=8-10$, for which the lattice artefact makes a small deviation. 

From Fig.~\ref{fig:noneqCP2}, a blow-up at early times, we see that because Higgs winding can only take place in the presence of a local Higgs field zero, the asymmetry is created when the average Higgs field is low in its oscillation. When it is high, winding number is essentially constant. Also, since the C(P)-violating force is proportional to $\dot{|\phi^2|}$, the gauge field picks up speed in-between Higgs extrema. The net effect is a ``pumping'' behaviour, of the baryon asymmetry as Higgs symmetry breaking proceeds. As the Higgs approaches a uniform vev, and oscillations damp out, the asymmetry creation gradually stops. This is very similar to the case in the 3+1 dimensional SU(2)-Higgs model \cite{Tranberg:2003gi,Tranberg:2006dg,Tranberg:2006ip}, where the gauge field is much more complicated dynamically than the model considered here. 

\DOUBLEFIGURE{./Pictures/fermback.eps,width=7cm,clip}{./Pictures/kappadep.eps,width=7cm,clip}
{The evolution of average Chern-Simons number, with (black) and without (red) fermion back-reaction. Here shown for $v=64$ (small effect) and $v=8$ (some effect).\label{fig:noneqCP3} }{Dependence of the asymmetry on C(P)-violation $\kappa$, measured at $m_At=150$. The behaviour is linear for small $\kappa$. Note that the asymmetry is zero at $\kappa=0$ by construction.\label{fig:noneqCP4} }

One important question is to what extent fermions can be ignored dynamically, compared to the bosonic fields. If not, many simulations of baryogenesis may need corrections from the fermion backreaction. In Fig.~\ref{fig:noneqCP3} we show the average Chern-Simons and fermion number with and without fermion backreaction, for $v=64$ and $v=8$, respectively. Clearly, for $v=64$, fermion backreaction can be mostly ignored, and the fermions only serve as spectator fields, encoding the fermion asymmetry. For this case, one might as well just do bosonic simulations, and infer fermion number from the anomaly equation a posteriori. For $v=8$ however, the fermions begin to influence the evolution of the gauge field, even when the bosonic fields are subject to a tachyonic instability and therefore grow large. For the Standard Model in 3+1 dimensions, there is no $v$-ambiguity, and it will be crucial, to what extent back-reaction is important. In particular, CP-violation itself is a backreaction effect, which will dynamically generate effective terms similar to the bosonic C(P)-violation used here. 

Finally, to illustrate the type of calculations that are possible, we show how the asymmetry depends on $\kappa$ (Fig.~\ref{fig:noneqCP4}). For small enough $\kappa$, the dependence is nicely linear. Also remember that because our scalar ensemble is explicitly $C$-symmetric, $\langle N_f(\kappa=0)\rangle=0$.

\section{Conclusion\label{sec:conc}}

By combining the methods of \cite{Aarts:1998td} and \cite{Borsanyi:2008eu}, we have demonstrated how to do first-principle numerical simulations of bosonic scalar-gauge systems with quantum fermions, in a numerically efficient manner. Although there is no gain in numerical effort in the specific 1+1 dimensional toy model considered here (compared to \cite{Aarts:1998td}), in the physically relevant case of 3+1 dimensions, we expect a significant decrease in the required computing time. As an example, including fermions in the simulations of \cite{Tranberg:2003gi}, an ensemble of $N_q=2430$ should be compared to the $90^3(=729,000)$ mode functions otherwise required, a gain-factor of $300$. Either way, fermions are numerically challenging, but for the setup of \cite{Tranberg:2003gi}, there would then be no need for the resource-consuming CP-violating term. Except for the scalar ensemble averages in section \ref{sec:CEB}, all the simulations presented here were done on a normal desktop computer in less than 24 hours in total. 

We found that Gauss' law, fermion back-reaction as well as the baryon anomaly are well reproduced in terms of a statistical ensemble of $\mathcal{O}(1000)$ fermion field realisations. As described in \cite{Borsanyi:2008eu} implementation of the fermion correlators, including the anti-commutativity of the fermionic operators requires a doubling of the fields into ``male'' and ``female'' (not to be confused with the standard lattice fermion doublers), adapted to the system at hand. In addition, the usual lattice doubler problem has to be addressed; in the present case we found that Wilson fermions in space and a small timestep was sufficient to keep the doublers sufficiently decoupled that they stayed un-excited for the timescales required here. Failure to do this leads to an exact cancelling out of the baryon anomaly. 

The method requires careful consideration of the interplay between lattice size, ensemble size and the size of couplings. In particular, the Yukawa coupling introduces additional lattice artefacts, which have to be compensated for, and we have demonstrated how to do this.

The upshot is that fermions are included completely in the dynamics, since they are bi-linear in the action, and so at least in cases where gauge fields are dominated by large particle numbers and long wavelength, this approach provides a very reliable description of the full field dynamics. The obvious application of the method is (electroweak) baryogenesis, where baryon number violating processes are classical in nature, whereas the CP-violation\footnote{At least in the Standard Model.} and the actual baryon number are carried by the fermion degrees of freedom. This applies both to ``hot'' and ``cold'' baryogenesis.

Simulations of the early stages of the heavy-ion collisions are also within the scope of the work presented here, since it involves very large (boosted) gluons fields coupled to (sea and valence) quarks. The valence quarks source the gauge field, which then evolves and may in turn source the emission of fermions. With the method presented here, fermions may be included in the dynamics completely. 

The obvious next step is to implement ensemble fermions in 3+1 dimensions, coupled to SU(2)-Higgs bosonic fields as in the Standard Model, where the gauge-fermion interaction is chiral rather than axial. Including the Standard Model CP-violation via the CKM matrix will require all three fermion generations, and represents a significant numerical challenge; with the method described here, this numerical effort can be reduced by one or even two orders of magnitude. This is work in progress.

\appendix

\section{Conventions}
\label{app:cons}
We use the metric signature $(-,+)$, and for the Dirac algebra, we employ the Weyl-Majorana representation
\ba
\{\gamma^\mu,\gamma^\nu\}=2\eta^{\mu\nu},\quad \bar\psi=i\psi^\dagger\gamma^0,\quad \gamma_5=-\gamma^0\gamma_1,
\ea
with explicitly
\ba
\gamma^0&=&-i\sigma^2=
\left(\begin{array}{cc}0  & -1\\1 & 0\end{array}\right)
,\quad
\gamma^1=\sigma^1=
\left(\begin{array}{cc}0 & 1\\1 & 0\end{array}\right)
,\quad
\gamma_5=-\gamma^0\gamma^1=
\left(\begin{array}{cc}1 & 0\\0 & -1\end{array}\right)
,\\
\mathcal{C}&=&\left(\begin{array}{cc}0 & -i\\i & 0\end{array}\right)
,\quad
\gamma^{\mu T}=-\mathcal{C}\gamma^\mu \mathcal{C}^{-1}.
\ea

\section{Lattice equations}
\label{app:lattice}

On the lattice, we define the gauge link variables ($a_\mu$, $\mu=0,1$ are the lattice spacings),
\ba
U_\mu(x)=\exp(-ia_\mu A_\mu(x)),\quad U^q_\mu(x)=\exp(-iqa_\mu A_\mu(x)).
\ea
We are using the non-compact formulation of the gauge action, with $A_\mu$ the basic gauge field variable. We define the derivatives
\ba
\label{eq:latticeDerivs}
\partial_\mu A_\nu(x) = \frac{1}{a_\mu}\left(A_\nu(x+\mu)-A_\nu(x)\right),&\quad& \tilde D_\mu=\frac{1}{2}\left[D_\mu+D'_\mu\right],\\
D_\mu\phi=\frac{1}{a_\mu}[U_\mu(x)\phi(x+\mu)-\phi(x)],&\quad&
D'_\mu\phi=\frac{1}{a_\mu}[\phi(x)-U^\dagger_\mu(x-\mu)\phi(x-\mu)],\\
D_\mu\psi(x)=\frac{1}{a_\mu}[U^q_\mu(x)\psi(x+\mu)-\psi(x)],&\quad& D'_\mu\psi(x)=\frac{1}{a_\mu}[\psi(x)-U^{q\dagger}_\mu(x-\mu)\psi(x-\mu)],\nonumber\\
\ea
We will deal with the spatial fermion doublers by including a Wilson term
\ba
W_1\psi&=&-\half r_1a_1 D'_1 D^1\psi.
\ea
The lattice action then becomes, 
\ba
S_{\rm Lat}= S_H+S_A+S_{F},
\ea
with
\ba
S_H&=&\sum_{x,t} a_1a_0\left[D_0\phi^\dagger D_0\phi-D_1\phi^\dagger D_1\phi-\lambda(\phi^\dagger\phi-v^2/2)^2\right],\\
S_A&=&\sum_x \frac{a_0a_1}{2e^2}\left(\partial_0A_1(x)-\partial_1 A_0(x)\right)^2,\\
S_{F}&=&-\sum_{x,t}a_0a_1\left[\bar{\psi}\left(\half\gamma^\mu (D_\mu+D_\mu')+W\right)\psi+\half G\phi^*\psi^T\mathcal{C}^\dagger\psi-\half G\phi\bar{\psi}\mathcal{C}\bar{\psi}^T\right].
\ea
This immediately gives the lattice equations of motion
\ba
D_\mu D'^\mu\phi-2\lambda(\phi^\star\phi-v^2/2)\phi-\frac{G}{2}\psi^TC\psi&=&0,\\
\label{eq:lattFermi}\gamma^\mu \tilde D_\mu\psi-\frac{a_1r_1}{2}D_1D_1'\psi+G\phi\psi^\star&=&0,\\
\del_\mu (\del'^\mu A^\nu-\del'^\nu A^\mu)+e^2(j_f^\nu+j_b^\nu+j_W^\nu)&=&0,
\ea
and the currents,
\ba
j_{b,\mu}&=&i(\phi D_\mu \phi^\star-\phi^\star D_\mu\phi),\\
\label{eq:latticeFermiCurrent1}
j_{f,\mu}&=&\frac{iq}{2}\left[\bar\psi(x)\gamma_\mu U^q_\mu(x)\psi(x+\mu)+\bar\psi(x+\mu)\gamma_\mu U^{q\star}_\mu(x)\psi(x)\right],\\
j_W^0&=&0,\\
j_W^i&=&iq\frac{a_1r_1}{2}\left[D^a\bar\psi-\bar\psi D^i\psi\right],
\ea
and so we have Gauss' law from
\ba
\del'_\mu (j_b^\mu+j_f^\mu+j_W^\mu)&=&0.
\ea
The chiral current is
\ba
j_{\mu,5}&=&\frac{i}{2}\left[\bar\psi(x)\gamma_\mu\gamma_5U_\mu(x)\psi(x+\mu)+\bar\psi(x+\mu)\gamma_\mu\gamma_5U_\mu^\star(x)\psi(x)\right],\\
\del'_\mu j^\mu_5&=&0.
\ea
Chern-Simons number and fermion number is trivially adapted to the lattice, and following \cite{Kajantie:1998bg}, we write $\phi(x)=|\phi(x)|e^{i\theta(x)}$, and then define the integer lattice winding number as
\ba
N_W=\frac{1}{2\pi}\sum_x \left[\theta(x+1)-\theta(x)+A_1(x)\right]_\pi-A_1(x).
\ea

\section{Fermion doublers}
\label{app:doublers}

\DOUBLEFIGURE{./Pictures/doublers.eps,width=7cm,clip}{./Pictures/doublers2.eps,width=7cm,clip}
{Tachyonic transition run at $v=64$, $\kappa=0$, $N_{q}=270$ with different values of the Wilson coefficient $r_1$. Without the Wilson term, the spatial doublers cancel out the anomaly. $r_1=1$ seems a good choice. \label{fig:doublers} }{ The same tachyonic transition as on the left, now with different values for the timestep $dt$. Even at the largest timestep, the timelike doublers are not sufficiently excited to cancel out the anomaly. We use the smalest timestep shown here, $dt=0.05$.\label{fig:doublers2}}

Doublers contribute to the anomaly with the opposite sign to the non-doubler modes, and the anomaly will then average out if we do not remove the doublers from the dynamics. Fig.~\ref{fig:doublers} shows the anomaly in a tachyonic transition for ``naive'' fermions $r_1=0$ and with a Wilson term $r_1>0$. The anomaly disappears for the naive fermions, when doublers are allowed to get excited.

By adding a Wilson term in space, but not time, the fermion equation will still lead to temporal doublers, so where we thought we were evolving a single Fermi-field $\psi$, we are actually evolving two Fermi-fields, which we call $\psi^+$ and $\psi^-$. To see this we define
\ba
\psi(t,x)&=&\left\{
\begin{array}{cc}
\psi^+(t,x)-\gamma_1\psi^- & \qquad \textnormal{if t is even},\\
\psi^+(t,x)+\gamma_1\psi^- & \qquad \textnormal{if t is odd}.
\end{array}
\right.
\ea
Now, if the lattice time derivative is evaluated on an even $t$ slice, then it actually only uses fields evaluated on the preceeding odd $t$ slice, and the following odd $t$ slice, in which case one finds the equation of motion (\ref{eq:lattFermi}) becomes
\ba
0&=&                \left[\gamma^\mu\tilde D_\mu\psi^+(x)-\frac{a_1r_1}{2}D_1D'_1\psi^+(x)+G\phi(x)\psi^{+*}(x)\right]\\\nonumber
  &~&-\gamma_1\left[\gamma^\mu\tilde D_\mu\psi^-(x)-\frac{a_1r_1}{2}D_1D'_1\psi^-(x)+G\phi(x)\psi^{-*}(x)\right].
\ea
Similarly, on odd $t$ slices we have
\ba
0&=&                \left[\gamma^\mu\tilde D_\mu\psi^+(x)-\frac{a_1r_1}{2}D_1D'_1\psi^+(x)+G\phi(x)\psi^{+*}(x)\right]\\\nonumber
  &~&+\gamma_1\left[\gamma^\mu\tilde D_\mu\psi^-(x)-\frac{a_1r_1}{2}D_1D'_1\psi^-(x)+G\phi(x)\psi^{-*}(x)\right],
\ea
showing that both $\psi^+$ and $\psi^-$ satisfy the fermion equation of motion, and that we are actually evolving two fermi degrees of freedom. 

In Fig.~\ref{fig:doublers2} we see a set of runs where we vary the timestep. We only initialise the non-doubler modes, and we see that for all timesteps that give a stable integration of the equations of motion, the time-like doublers stay un-excited. In all simulations in the main paper, we use $dt=0.05$, the smallest time-step presented here, and we see no doubler effects.

\section{Spinors \label{sec:spinors}}

We now construct the basis spinors required in the mode expansion of the fermion operators, which we do by setting $A_1=0$ and taking $|\phi|^2=v^2/2$ so that for $\Psi^1$, $m_f=Gv/\sqrt{2}$. For the positive frequency modes
\ba
\Psi_1&=&U_{1,k}e^{ik.x},\quad k^0=\omega_1>0,
\ea
we are then led to
\ba
U_{1,k}&=&
\left(
\begin{array}{c}
-i\sqrt{\omega_1+s_k} \\
\frac{M_1}{\sqrt{\omega_1+s_k}}
\end{array}
\right),\\
s_k&=&\frac{1}{a_1}\sin(a_1k),\quad m_k=\frac{r_1}{a_1}[1-\cos(a_1k)],\\
M_1&=&m_k+m_f,\qquad \omega_1=+\sqrt{(M_1)^2+(s_k)^2},
\ea
Similarly, the negative frequency solutions are found by
\ba
\Psi_1=V_{1,k}e^{-ik.x},\quad k^0=\omega_1>0,
\ea
but because the field is Majorana, and therefore real, we immediately have $V=U^*$. In order to calculate the two-point function we need the following identity,
\ba
U_{1,k}\bar U_{1,k}=M_1-i\gamma^\mu\tilde k_\mu,
\ea
where $\tilde k^\mu=(\omega_1,s_k)$. 
To get the mode functions for $\Psi^2$, simply make the replacement $m_f\rightarrow-m_f$.


\end{document}